\documentclass[12pt]{article}
\pdfoutput=1
\usepackage[title,titletoc,toc]{appendix}
\usepackage{hyperref}
\usepackage{pstricks}
\usepackage{color}
\usepackage{amssymb,amsmath,amsfonts,bm,bbm}
\usepackage{afterpage}
\usepackage{longtable}
\usepackage{cite}
\usepackage{latexsym, mathrsfs}
\usepackage{graphicx,graphics,epsf,epsfig,epstopdf,subfigure,psfrag}
\usepackage{url}
\usepackage{paralist}
\usepackage{bbold}
\newcommand{\be}{\begin{equation}}
\newcommand{\ee}{\end{equation}}
\newcommand{\bea}{\begin{eqnarray}}
\newcommand{\eea}{\end{eqnarray}}
\newcommand{\bec}{\begin{center}}
\newcommand{\eec}{\end{center}}
\newcommand{\nn}{\nonumber}

\newcommand{\dd}{\displaystyle}

\begin{document}

\begin{flushright}{BARI-TH/2016-701} \end{flushright}

\thispagestyle{empty}
\bigskip

\begin{center}
{\LARGE\bf On Exclusive $h \to V \ell^+ \ell^-$ Decays}\\[1.0 cm]
{\bf  Pietro~Colangelo$^{a}$,  Fulvia~De~Fazio$^{a}$, Pietro~Santorelli$^{b,c}$\\[0.5 cm] }
{\small
$^a$INFN, Sezione di Bari, via Orabona 4, I-70126 Bari, Italy\\
$^b$Dipartimento di Fisica "Ettore Pancini", Universit\`a  di Napoli Federico II, Complesso Universitario di Monte Sant'Angelo,
Via Cintia, Edificio 6, I-80126 Napoli, Italy\\
$^c$INFN, Sezione di Napoli, I-80126 Napoli, Italy\\
}
\end{center}

\bigskip

\abstract
We study a set of exclusive decay modes of the Standard Model Higgs boson into a vector meson and a dilepton pair:
$h\to V \ell^+ \ell^-$, with $V=\Upsilon, J/\psi,\phi$, and $\ell=\mu, \tau$,  determining the decay rates, the dilepton mass spectra and the $V$  longitudinal helicity fraction distributions.  In the same framework, we analyze the   exclusive  modes into neutrino pairs $h\to V \nu \bar \nu$. We also discuss the implications of the recent CMS and ATLAS results for the lepton flavor-changing process  $h\to \tau^+ \mu^-$ on the 
  $h\to V \tau^+ \mu^-$ decay modes.
\vspace*{1.0cm}


\vspace*{1cm}


Precision tests of the properties of the Higgs-like scalar with $m_h=125.7(4)$  GeV observed at the LHC  \cite{Aad:2012tfa,Chatrchyan:2012xdj,Agashe:2014kda}, to verify that the Standard Model (SM) predictions for  the Higgs boson are exactly fulfilled, represent an issue of prime interest in present-day theoretical and experimental activity. Particularly important is to confirm that the couplings of the observed state to the fermions and gauge bosons are what the SM dictates. 
The LHC measurements are consistent with the Standard Model predictions for the Higgs
couplings  to top and beauty quarks  and to $\tau$ leptons \cite{atlascms}, while  the couplings to the other quarks and  leptons are experimentally less known. Approaches based on 
 the effective field theory which includes  dimension $6$ operators show how such couplings could be modified,  comprising also  CP violating terms 
\cite{Contino:2013kra,Brivio:2013pma,Gonzalez-Alonso:2014eva,Gupta:2014rxa,Chien:2015xha}. In addition,
possible beyond SM lepton and quark flavor-changing Higgs couplings need to be examined.  This  issue  is  important in connection with the  current $h \to \tau \mu$ searches at LHC:  for such a  mode the CMS Collaboration has reported 
${\cal B}(h \to \tau \mu)=\left(0.84^{+0.39}_{-0.37}\right)\times 10^{-2}$ and the upper bound   ${\cal B}(h \to \tau \mu)<1.51 \times 10^{-2}$ at $95\%$ CL \cite{Khachatryan:2015kon}, while the ATLAS
Collaboration  quotes the bound  ${\cal B}(h \to \tau \mu)<1.85\times 10^{-2}$ at $95\%$ CL
\cite{Aad:2015gha}.

Measuring the Higgs couplings to the first two generation fermions is  a difficult task. Various possibilities have been studied,   with particular attention to the radiative  $h \to f \bar f \gamma$ processes.   The leptonic modes $h \to \ell^+ \ell^- \gamma$ (with $\ell=e, \mu$) have been considered in \cite{Abbasabadi:1996ze,Chen:2012ju,Sun:2013rqa,Dicus:2013ycd} and  \cite{Passarino:2013nka}. 
To access the Higgs couplings to the light quarks,  the exclusive channels $h \to V \gamma$, with $V$ a vector meson, have been scrutinized in \cite{Bodwin:2013gca, Kagan:2014ila,Isidori:2013cla,Koenig:2015pha}, 
and $h \to V Z$ have been studied in \cite{Bhattacharya:2014rra,Gao:2014xlv}.
Here, we examine the three-body exclusive Higgs decays 
$h \to V \ell^+ \ell^-$, where $V=\Upsilon, J/\psi, \phi$ and $\ell$ is a  light or a heavy charged lepton.  
 There are several motivations to afford such a study. The first one is the possibility of considering, in addition to the decay rates, some distributions encoding important physical information, namely the distributions in the dilepton invariant mass squared: we shall see, for example,  that  the case of  $\tau$  dileptons is particularly interesting. Moreover, 
since several amplitudes contribute to each process, one can look at kinematical configurations where  the interferences are more effective, in the attempt of getting information on the various Higgs couplings.   Deviations from the Standard Model can also be probed through the search of lepton flavor violating signals.  All the considered modes have a clear experimental signature, although the rates are small,  and can be included in the physics programme of  future high luminosity facilities.

\vspace*{0.5cm}
\begin{figure}[t!]
\begin{tabular}{ccc}
\includegraphics[width = .33\textwidth]{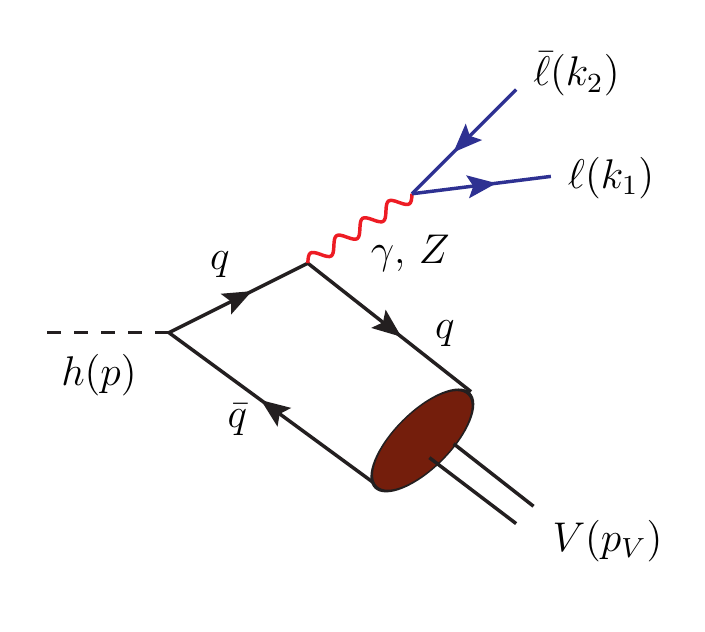} &
\includegraphics[width = .33\textwidth]{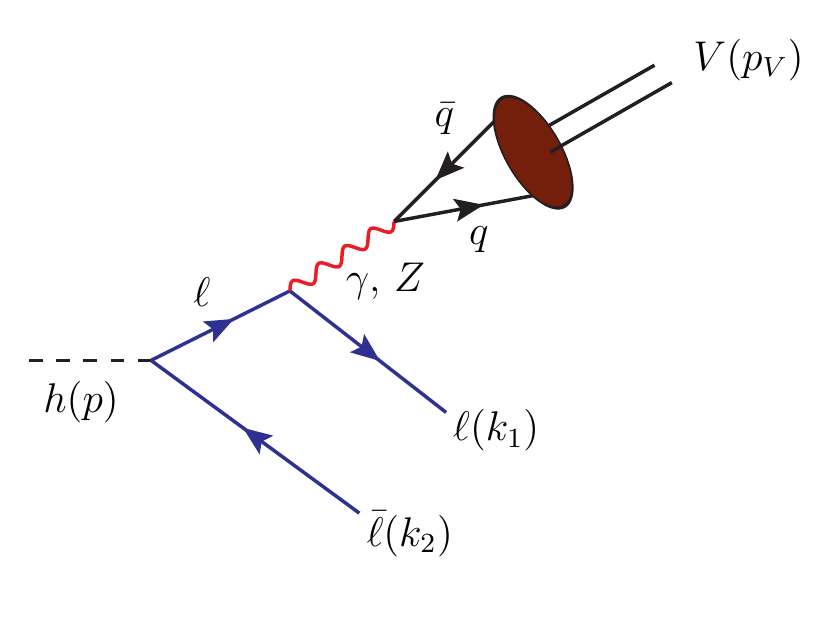} &
\includegraphics[width = .33\textwidth]{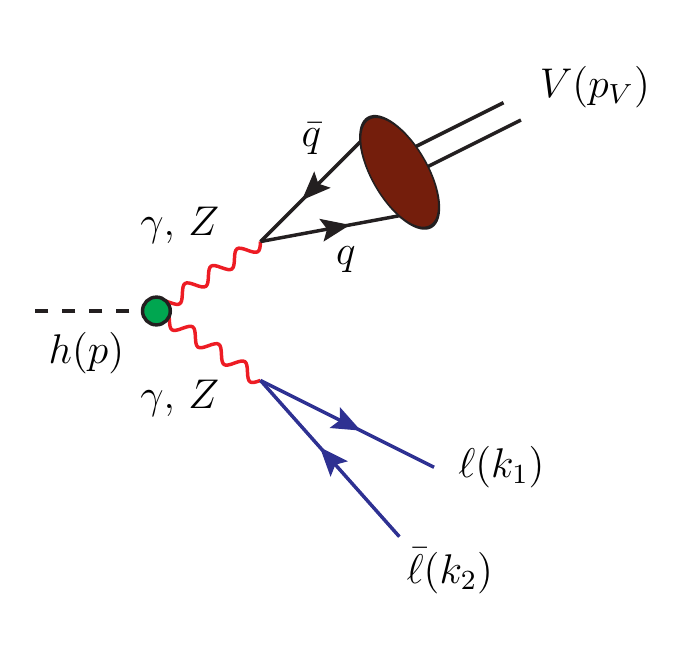}\\
(a)  &  (b) & (c) \\
\end{tabular}
\caption{\baselineskip 10pt Diagrams involved in $h \to V \ell^+ \ell^-$ decays. In  (a) and (b) the diagrams with $\gamma$ and $Z$ emitted by $\bar q$ and $\bar \ell$ are also considered. Diagram  (c) represents  the contribution of the $h \to ZZ$  vertex and of the effective $h \to \gamma \gamma$,  $h \to \gamma Z$ vertices.}\label{fig:diagrams}
\end{figure}

The decays $h \to V \ell^+  \ell^-$ take contribution from amplitudes in which the Higgs couples to quarks, to leptons and to the gauge bosons  $Z$ and $\gamma$.  In  SM such couplings are $g_{h f\bar f}=i \displaystyle{\frac{m_f}{v}}$ for fermions\footnote{For the quarks we use  the running masses evaluated at the Higgs mass scale $\mu\simeq m_h=125$ GeV at NNLO in the ${\overline {\rm MS}}$ scheme.},  and $g_{hZZ}=i \displaystyle{\frac{2m_Z^2}{v}}$  for  $Z$ ($v=2 m_W/g=(\sqrt 2 G_F)^{-1/2} = 246$ GeV 
is the Higgs field vacuum expectation value). The effective $\gamma \gamma$  and $Z \gamma$ Higgs  couplings are  described below.
Fig.~\ref{fig:diagrams} displays the three kinds of diagrams  that must be taken into account. 

Diagrams (a)  represent amplitudes with  the Higgs coupled to the quark-antiquark pair. The neutral gauge boson $\gamma$ or $Z$ is emitted from the quark or the antiquark before they hadronize in the vector meson $V$. For low dilepton invariant mass squared $q^2=(k_1+k_2)^2$, the nonperturbative quark hadronization in the vector meson $V$ can be analyzed adopting the formalism of the QCD hard exclusive processes \cite{Lepage:1979zb,Lepage:1980fj,Efremov:1979qk,Chernyak:1983ej}. The matrix elements of the non-local  quark-antiquark operator,  depicted in Fig.~\ref{fig:diagrams}(a),
and the vector meson can be expressed as  an expansion in increasing twists, which  involves various  vector meson distribution amplitudes. For   $V=\Upsilon,\,J/\psi,\,\phi$, the leading twist
   light-cone distribution amplitude (LCDA)  $\phi_\perp^V$  is defined from the matrix element of the non-local   ${\bar q}(y) \sigma_{\mu \nu} q(x)$ quark current:
   \be
\langle V(p_V ,\,\epsilon_V) | {\bar q}(y) \sigma_{\mu \nu} q(x)| 0 \rangle =-f_V^\perp (\epsilon_{V \, \mu}^* p_{V \, \nu} -\epsilon_{V \, \nu}^* p_{V \, \mu}) \int_0^ 1 du \, e^{i\,u\,p_V \cdot x+i\,{\bar u}\,p_V \cdot y} \phi_\perp^V(u)
\label{LCDA}
\ee
$({\bar u}=1-u)$.  $u\,p_V$  and ${\bar u}\,p_V$  represent the meson longitudinal momentum fraction  carried by the quark and antiquark.  $\phi_\perp^V$ is normalized to $1$; the hadronic parameter $f_V^\perp$ is discussed below. Tle LCDA  $\phi_\perp^V$ can be expressed in terms of the Gegenbauer polynomials $C_n^{3/2}$,
\be
\phi_\perp^V(u,\,\mu)=6u{\bar u}\left[1+\sum_{n=1}^\infty a_n^{V_\perp} (\mu) C_n^{3/2}(2u-1) \right] \,\,\, , \label{phiperp}
\ee
 with the scale $\mu$  dependence of the distribution amplitude encoded in the coefficients $a_n^{V_\perp} (\mu)$. Such coefficients follow
a renormalization group evolution
\be
a_n^{V_\perp} (\mu) = \left[ \frac{\alpha_s(\mu)}{\alpha_s(\mu_0)} \right]^{\gamma_n^\perp /(2 \beta_0)} a_n^{V_\perp} (\mu_0) \,\,\,\, , \label{evolution}
\ee
where $\gamma_n^\perp=8C_F \left(\sum_{k=1}^{n+1} \displaystyle{\frac{1}{k}} -1 \right)$, $C_F=\frac{N_c^2-1}{2 N_c}$ and $N_c$ the number of colors. We set the low-energy scale  $\mu_0 \simeq 1$ GeV.

It is convenient to distinguish between the  heavy  $J/\psi, \Upsilon$ and light $\phi$ mesons.
In the case of $\phi$, the expansion \eqref{phiperp} (where only the even momenta are non-vanishing) is known up to  $n=4$, with  values of the coefficients  \cite{Ball:2006eu,Ball:2007rt,Grossmann:2015lea} 
\be
a_2^{\phi_\perp}(\mu_0)= 0.14 \pm 0.07 \hskip 1cm ,  \hskip 1cm  a_4^{\phi_\perp}(\mu_0)= 0. 00\pm 0.15\,\,.
\ee
For heavy quarkonia  $V=J/\psi$ and $\Upsilon$,  models for LCDA have been proposed. We use
the gaussian model \cite{Grossmann:2015lea}
\be
\phi_V^\perp(u,\mu_0)=N_\sigma \frac{4u{\bar u}}{\sqrt{2\pi}\sigma_V} \exp \left[-\frac{u-\frac{1}{2}}{2 \sigma_V^2} \right], 
\ee
with $N_\sigma$  a  normalization  constant  and the parameter $\sigma_V$, specific for each vector meson,  taking the values
\be
\sigma_{J/\psi}= 0.228 \pm 0.057\hskip 1cm ,  \hskip 1cm  \sigma_\Upsilon= 0.112 \pm 0.028 \,\,.
\label{sigmas}
\ee
The Gegenbauer momenta 
\be
a_n^{V_\perp} (\mu_0) =\frac{2(2n+3)}{3(n+1)(n+2)}\int_0^1du \,C_n^{3/2}(2u-1)  \phi_V^\perp(u,\mu_0)
\ee
 are  evolved using (\ref{evolution}) to determine the distribution amplitude $\phi_V^\perp$ at the scale $\mu \sim m_h$. For $J/\psi$ and $\Upsilon$ we include   $n=20$ terms in the Gegenbauer expansion. 

To assess the limit of applicability of the twist expansion,
 we investigate the hierarchy between the leading term  included in our calculation,  involving the twist 2 distribution $\phi_\perp^V$  in (\ref{phiperp}), and the next-to-leading term. 
This involves the combination $B(u)=h_\parallel^{(t)}(u)-\frac{1}{2} \phi_\perp^V(u)-\frac{1}{2}h_3(u)$ of  the distributions $h_\parallel^{(t)}$ and $h_3$ of twist 3 and 4, respectively \cite{Ball:1998ff}.
While the contribution of the leading term  contains the quark propagator $p_1=1/(m_h^2 {\bar u}+u q^2-u {\bar u}m_V^2)$, the next term involves $ p_2={m_V^2}/{\left(m_h^2 {\bar u}+u q^2-u {\bar u}m_V^2\right)^2}$;  in the case of the diagram with intermediate antiquark  $u \leftrightarrow {\bar u}$ should be exchanged.
The hierarchy  $p_1>p_2$   is always verified except  close to the  endpoint ${\bar u}=0$ (or $u=0$), where however the wave functions vanish.  Hence, the expansion can be trusted  up to quite large values of $q^2$.

A second issue is related to the role of ${\cal O}(\alpha_s)$ corrections from gluon exchanges among the quarks in the 
diagrams in Fig.~\ref{fig:diagrams}~(a) (they do not  need  to be included in the topologies when  the experimental values for the decay constants $f_V$ are used). The calculation of the corrections for $h \to V \gamma$ has been carried out  at $q^2=0$  for a real photon \cite{Koenig:2015pha}.  An  estimate of the size of the corrections in the cases considered here can be obtained extending the result  to  the whole $q^2$ range. The integrand functions in $I_1({\hat q}^2)$ and $I_2({\hat q}^2)$ in \eqref{int1},  \eqref{int2} are modified  as
\be
I_1^{\alpha_s}( {\hat q}^2)=\int_0^1\,du \, \phi_\perp^V(u)\left[\frac{1}{D_1(1-u,u,{\hat q}^2)}+\frac{1}{D_1(u, 1-u,{\hat q}^2)} \right]\ \left[1+\frac{C_F \alpha_s(\mu)}{4 \pi} h(u,m_h,\mu) \right]\,,\label{I1radcorr}
\ee
with \cite{Koenig:2015pha}
\be
h(u,m_h,\mu)=2 \ln[u(1-u)]\left(\log{\left( \frac{m_h^2}{\mu^2}\right)}-i \, \pi \right) +\ln^2(u)+\ln^2(1-u)-3 \,\,\,\, ,
\ee
and similarly for $I_2$.
The channels with  larger effects  are those with final $\Upsilon$,  where  ${\cal O}(\alpha_s)$ corrections  affect $I_{1,2}(\hat q^2)$ at $30\%$ level  close to  ${\hat q}^2 \simeq 0$,   and  decrease  when ${\hat q}^2$ is increased. 
The correction  modifies the results for the rates by about $10\%$, as $I_1$ and $I_2$ enter in the amplitude with opposite signs.
 
The Higgs couplings to leptons are involved in the diagrams  in Fig.~\ref{fig:diagrams}~(b), with the $q \bar q$  pair  emitted by the photon or  $Z$. Such diagrams  are  important  in the case of $\tau$. 
The hadronization of the $q \bar q$  pair into  the vector meson is described by the matrix element
\be
\langle  V(p_V ,\,\epsilon_V) | {\bar q}\, \gamma_\mu\,  q| 0 \rangle=-i f_V m_V \epsilon^*_{V \,\mu} \,\, ,
\label{fV}
\ee
with $p_V$ and $\epsilon_V$ the $V$ meson momentum and polarization vector, respectively.
The decay constant $f_V$ can be extracted from the $V \to e^+ e^-$ measured width. On the other hand,  the hadronic parameter $f_V^\perp$  in \eqref{LCDA} is less accessible, and results from lattice or QCD sum rule computations must be used.  In our analysis we use the  range for the ratio $R_{f_V}=\frac{f_V^\perp}{f_V}$ quoted in \cite{Grossmann:2015lea},   obtained exploiting  non-relativistic QCD  scaling relations \cite{Caswell:1985ui,Bodwin:1994jh}:
\bea
f_\phi &=&0.223 \pm 0.0014 \,\,\, {\rm MeV}\,\,\, ,  \hskip 1.5cm R_{f_\phi}=0.76 \pm 0.04 \,\,, \nn \\
f_{J/\psi} &=& 0.4033 \pm 0.0051 \,\,\, {\rm MeV}\,\,\, , \hskip 1cm R_{f_{J/\psi}}=0.91 \pm 0.14 \,\,, \label{constants} \\
f_\Upsilon &=&0.6844 \pm 0.0046 \,\,\, {\rm MeV}\,\,\, , \hskip 1.3cm R_{f_\Upsilon}=1.09 \pm 0.04 \,\,. \nn
\eea

The diagrams  in   Fig.~\ref{fig:diagrams}~(c)  involve the coupling of the Higgs  to a pair of gauge bosons, which in turn are coupled to a  lepton pair and to a $q \bar q$ pair  that hadronizes  into $V$.
The  elementary $hZZ$  coupling  can be read  from the SM Lagrangian. The effective $h\gamma \gamma$ and $hZ\gamma$   vertices  can be  written as
\be
A(H \to  G_1 G_2)=i \frac{\alpha}{\pi v}C_{G_1 G_2}\left[g_{\mu \nu}(p_V \cdot q)-p_{V\mu} q_\nu \right]  \epsilon_{G_1}^{*\mu} \,\, \epsilon_{G_2}^{*\nu}  \,\,\, , \label{effective}
\ee
with $G_1$ and $G_2$  either $\gamma \gamma$ or $Z\gamma$,  and $\epsilon_{G_1}$, $\epsilon_{G_2}$   polarization vectors.
In Eq.~(\ref{effective}) $p_V$ is the momentum  of the meson V and $q$ the momentum of the dilepton.  The effective  $h \gamma \gamma$ and $hZ\gamma$ couplings are determined by loop diagrams:  $C_{\gamma \gamma}=-3.266 +i 0.021$ and $C_{\gamma Z}=-2.046 +i 0.005$ \cite{Koenig:2015pha}. 
In the $Z$ propagator,  the width $\Gamma(Z)=2.4952$ GeV is included neglecting  its  small uncertainty \cite{Agashe:2014kda}. 
It is worth remarking that the possibility to access the $hZZ$ coupling is a feature of the class of modes we are analyzing. Moreover, since a sizeable contribution to $h \to V  \ell^+ \ell^-$  involves the effective  $h\gamma\gamma$ and $h Z \gamma$  couplings   from diagrams  sensitive to  New Physics effects,   the exclusive processes also probe  deviations from SM. 

The relative role of the diagrams in Fig.~\ref{fig:diagrams} is different if the  dilepton invariant mass is varied.
At  low-$ q^2$  the  amplitudes  with the Higgs coupled to the quarks provide a  contribution which decreases with $q^2$.  This contribution is sizable for  $\Upsilon \mu^+ \mu^-$.
Increasing $q^2$, the role of the other diagrams becomes important, and the uncertainty  in the terms   in Fig.~\ref{fig:diagrams}~(a) is overwhelmed by the other errors. 
 At large $q^2$ the contribution is also estimated to be smaller than the uncertainty affecting the other diagrams, as one can infer  modelling, e.g., the photon amplitude  with the inclusion of a set of intermediate states.

To compute the branching fractions, it is necessary to get rid of the poorly known Higgs full width. One possibility is to use  the expression  
\be
 {\cal B}(h \to V \ell^+ \ell^-) = \frac{\Gamma (h \to V \ell^+ \ell^-)}{\Gamma (h \to  \gamma \gamma)} {\cal B}(h \to \gamma \gamma)_{exp} \,\,\, 
 \ee 
which employs the computed  widths $\Gamma (h \to V \ell^+ \ell^-)$ and $\dd \Gamma (h \to \gamma \gamma)=\frac{\alpha^2}{64 \pi^3 v^2} |C_{\gamma \gamma}|^2 m_h^3$  combined with  the measurement  $\dd {\cal B}(h \to \gamma \gamma)_{exp}=(2.28 \pm 0.11 ) \times10^{-3}$ \cite{Heinemeyer:2013tqa}.  We obtain the following results:
\bea
{\cal B}(h \to \phi \mu^+ \mu^-)&=&(7.93 \pm 0.39)  \times 10^{-8}\nn \\
{\cal B}(h \to \phi \tau^+ \tau^-)&=&(2.35 \pm 0.12) \times 10^{-6}  \nn \\
{\cal B}(h \to J/\psi \mu^+ \mu^-)&=&(9.10 \pm 0.50) \times 10^{-8} \nn  \\
{\cal B}(h \to J/\psi \tau^+ \tau^-)&=&(1.82 \pm 0.10) \times 10^{-6}  \label{br} \\
{\cal B}(h \to \Upsilon \mu^+ \mu^-)&=&(5.60 \pm 0.37) \times 10^{-7} \nn \\
{\cal B}(h \to \Upsilon \tau^+ \tau^-)&=&(5.66 \pm 0.29) \times 10^{-7}\,\,\,\, . \nn
\eea

\begin{figure}[t!]
\includegraphics[width = .30\textwidth]{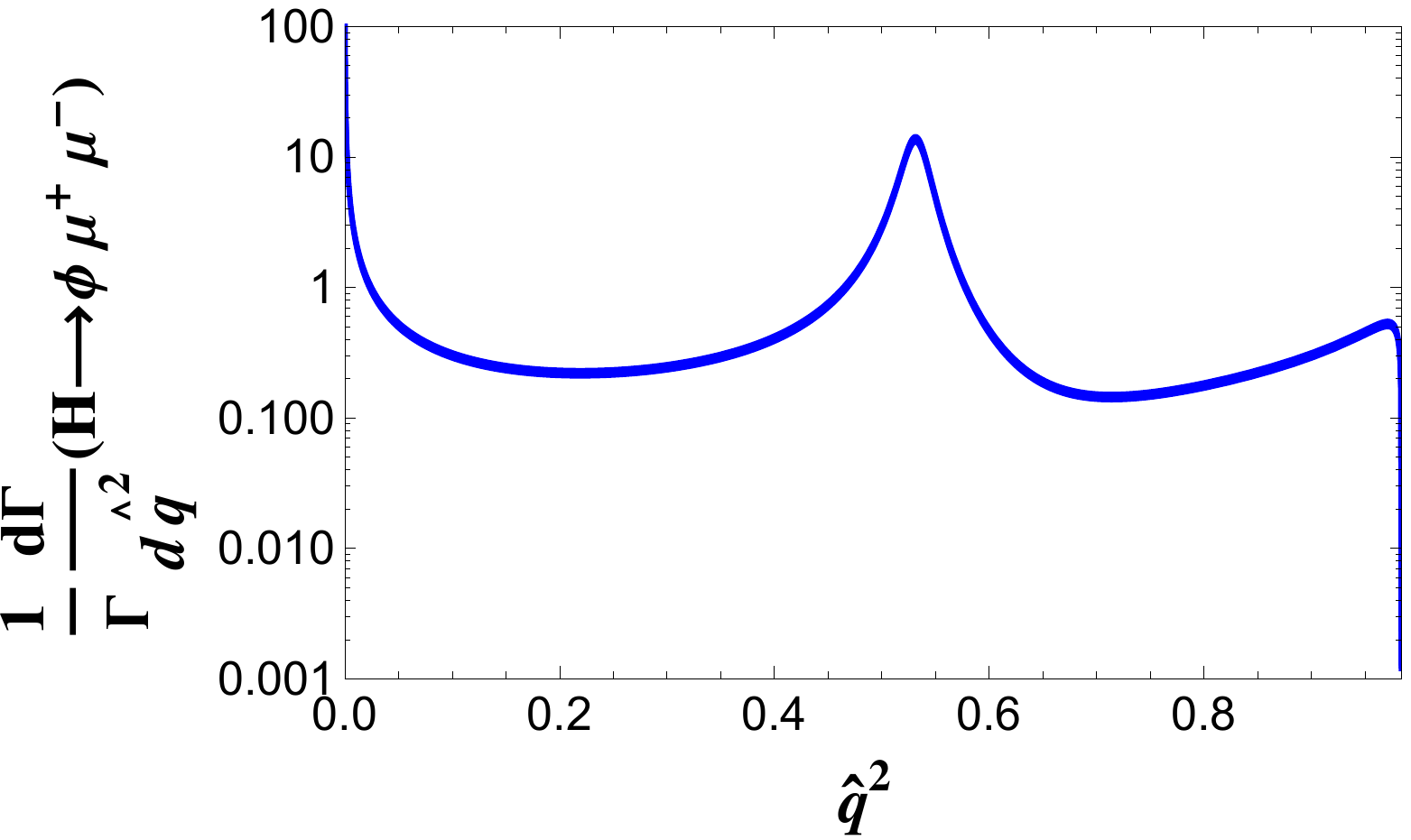}
\includegraphics[width = .30\textwidth]{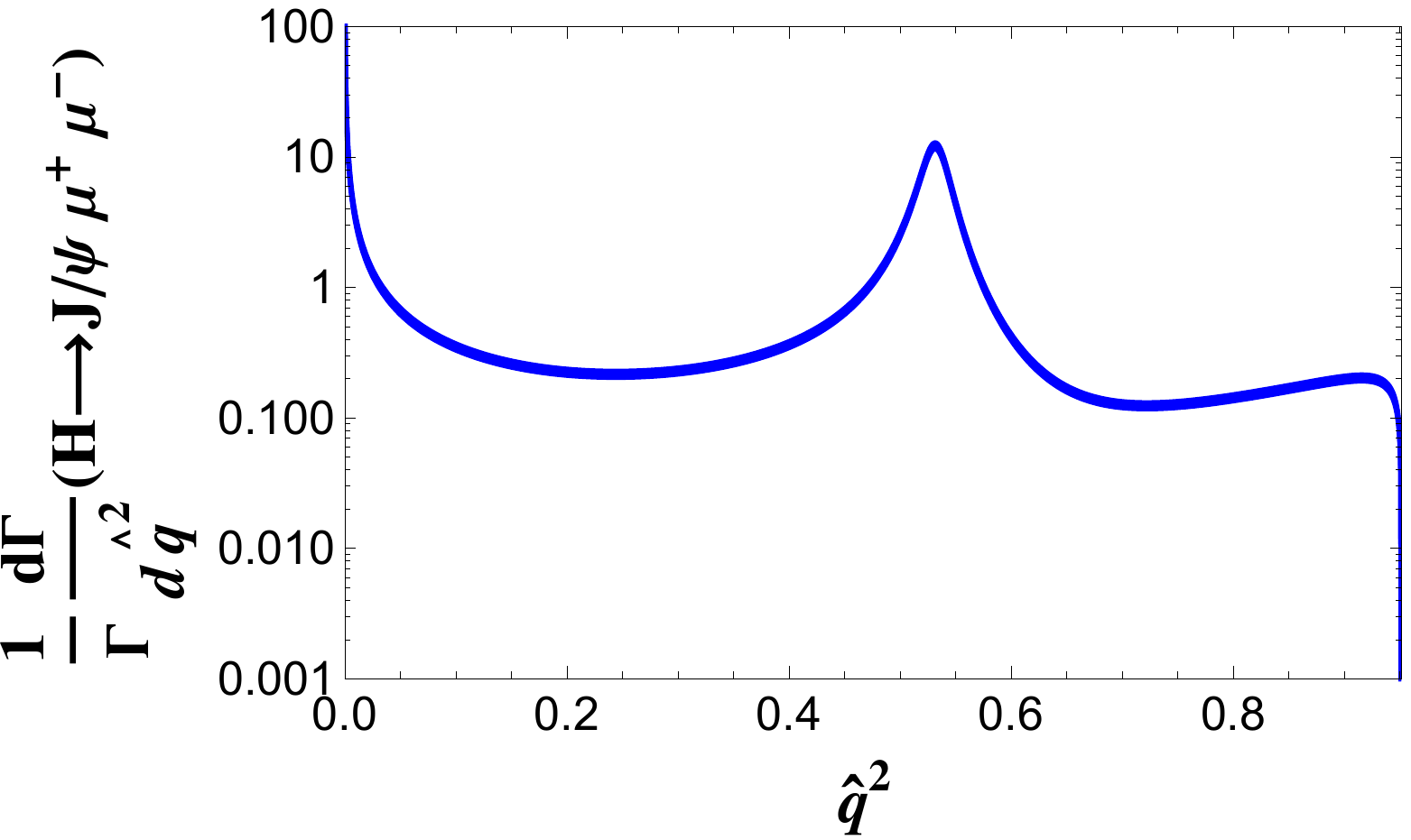}
\includegraphics[width = .30\textwidth]{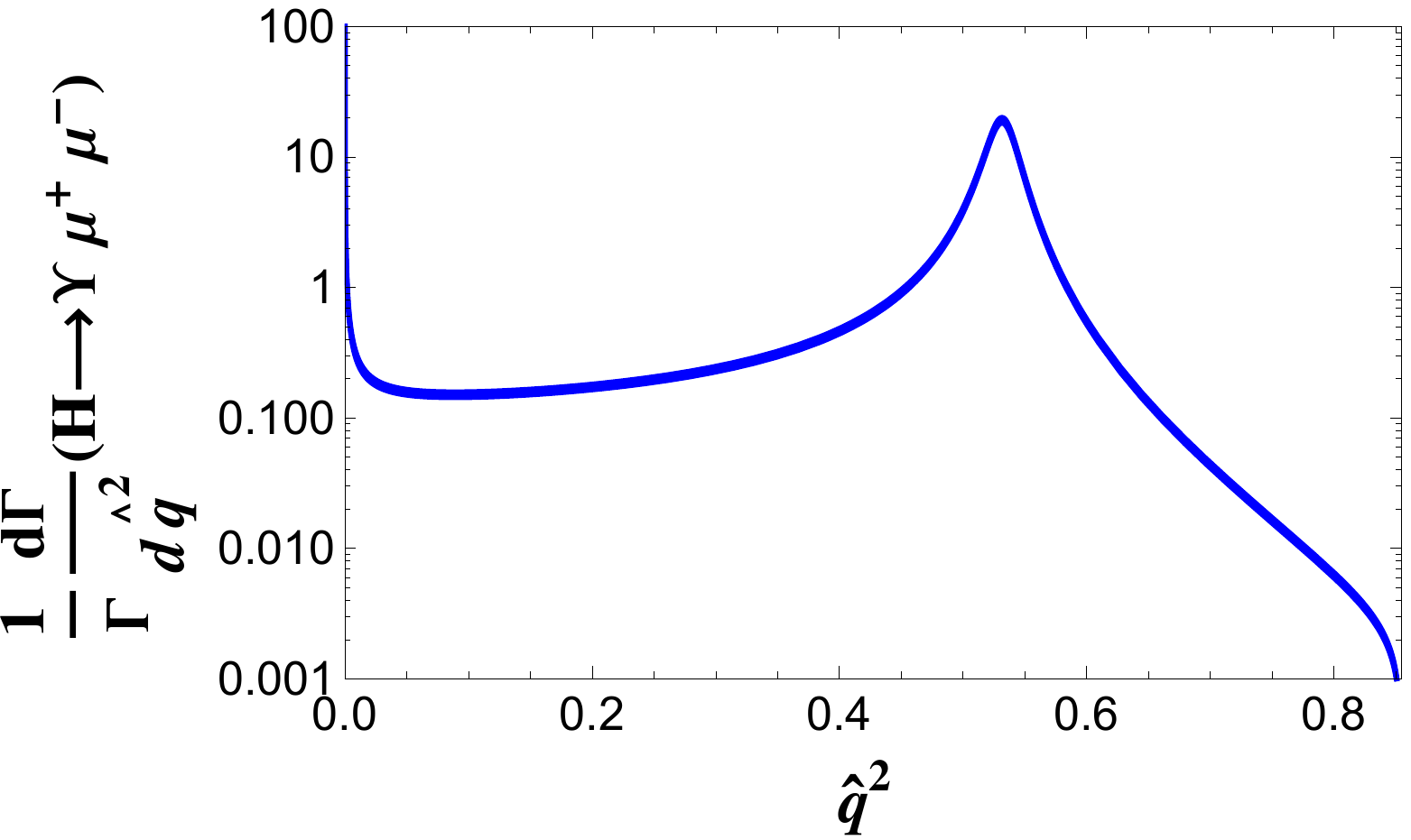}
\\
\\
\includegraphics[width = .30\textwidth]{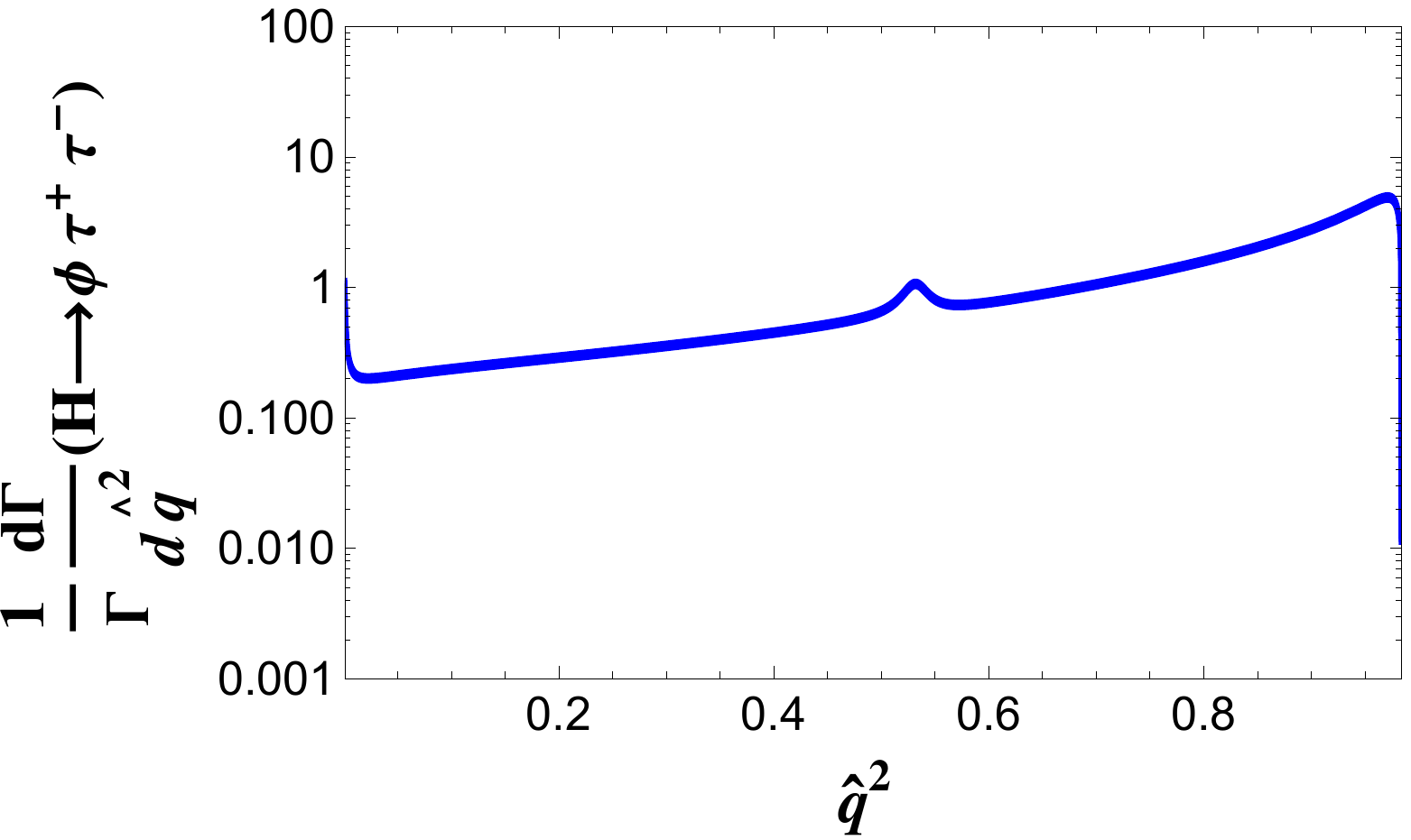}
\includegraphics[width = .30\textwidth]{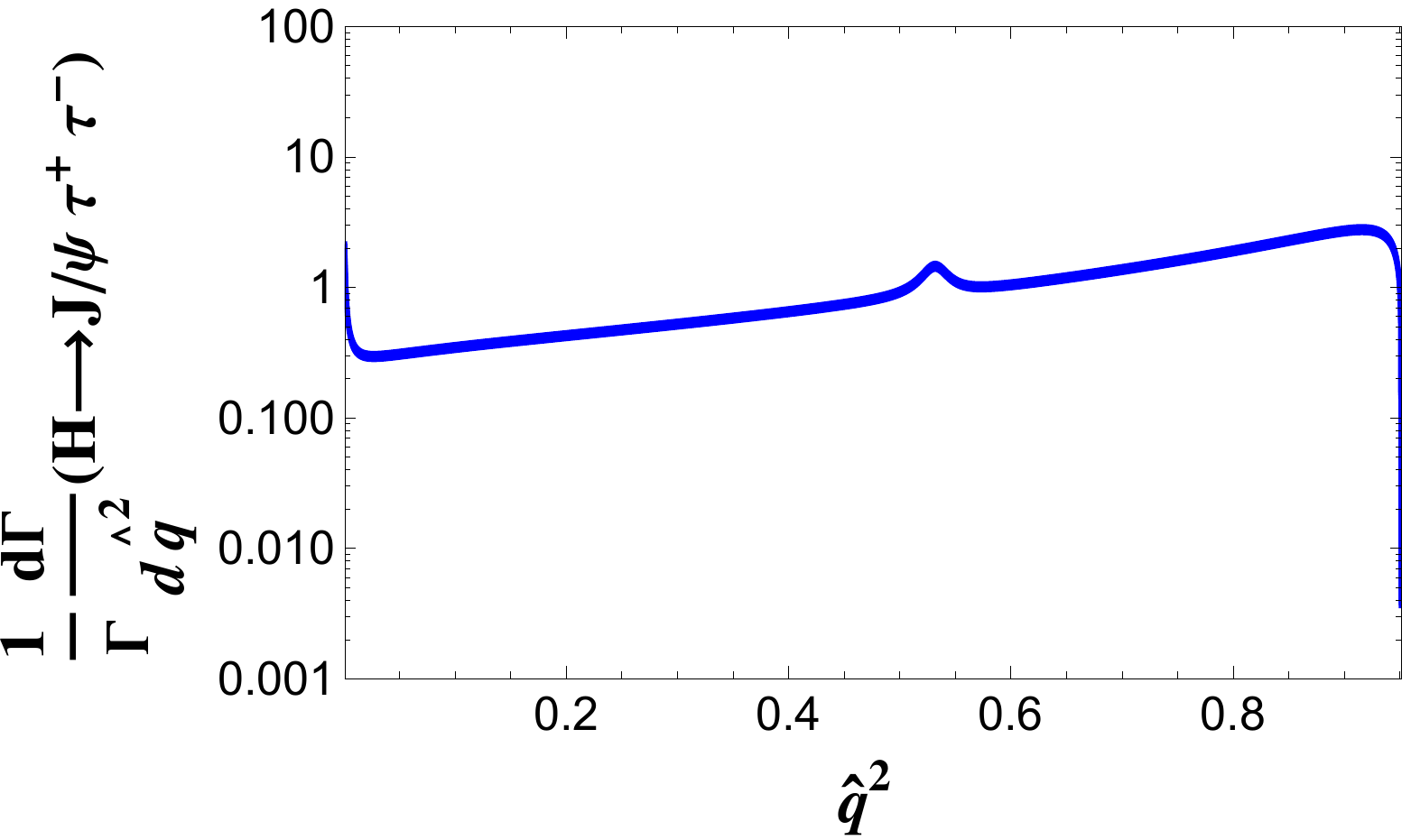}
\includegraphics[width = .30\textwidth]{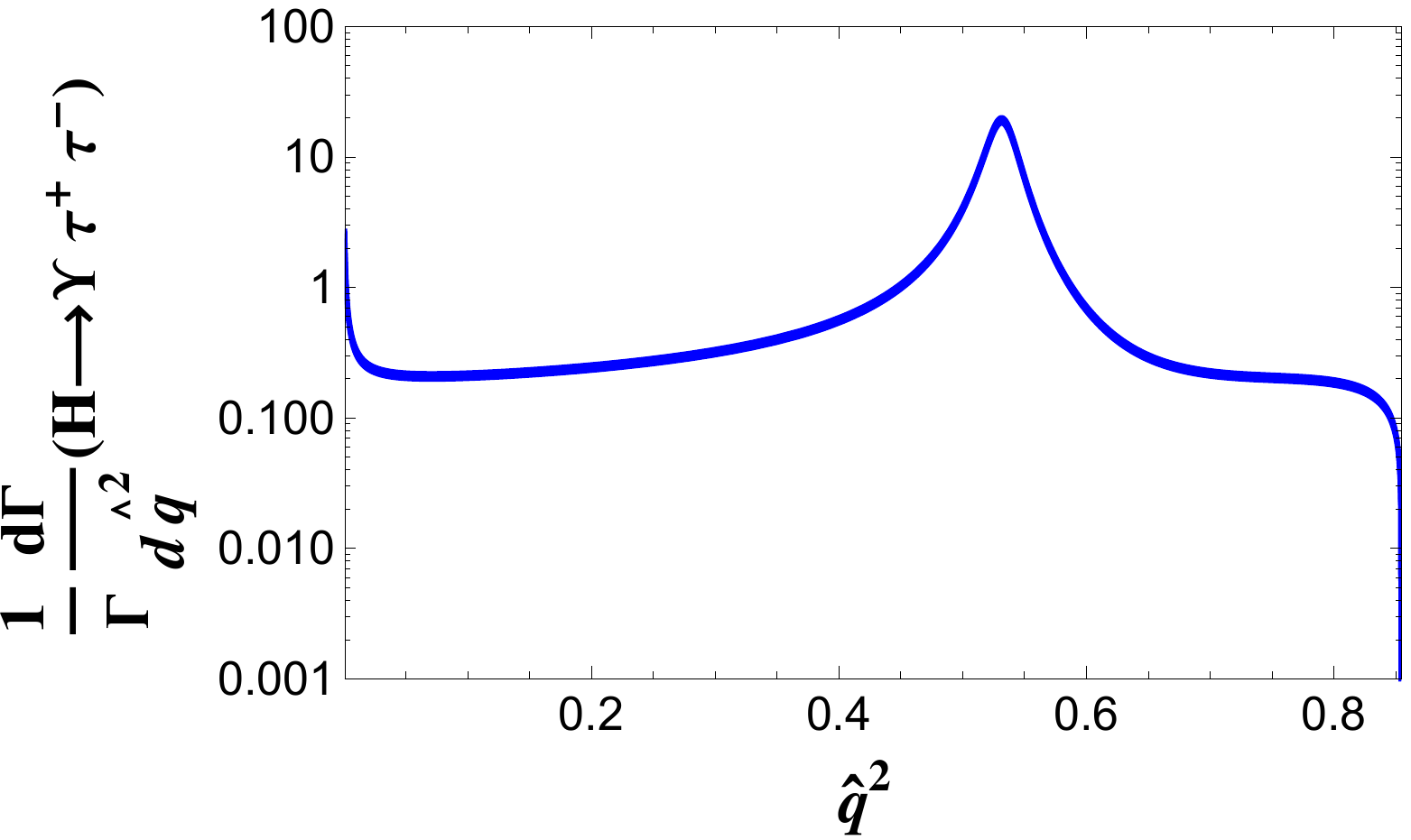}
\caption{Normalized decay distributions $\left(1/\Gamma \right) \, d {\Gamma}(h \to V \ell^+ \ell^-)/d \hat q^2$, with $\hat q^2= \frac{q^2}{m_h^2}$ and $q^2$ the dilepton mass squared.}\label{fig:distributions}
\end{figure}

The errors in  the  branching ratios  include the uncertainties on the LCDA parameters, on the decay constants $f_V$ and on the ratios $R_{f_V}$ in Eqs. (\ref{constants}), and the error on ${\cal B}(h \to \gamma \gamma)_{exp}$.
The   uncertainties on $f_V$ and   on  the meson LCDA  parameters give a small contribution to the errors in \eqref{br}, which are instead dominated by the uncertainty on  ${\cal B}(h \to \gamma \gamma)_{exp}$ amounting to $50-60\%$ of the total error, for the various channels.  The  uncertainty on $R_{f_V}$  constitutes $20-30\%$ of the total error. The uncertainty from the $\alpha_s$ corrections is not included in the error budget.

The larger rates in \eqref{br} are predicted  for  modes with $\tau$ pairs,  $h \to \phi \tau^+  \tau^-$ and $h \to J/\psi \tau^+  \tau^-$. The  modes with muons have rates suppressed by a factor 30 and 20, respectively,    that could be experimentally overcome by the  identification efficiency. In the case of $\Upsilon$, the modes with $\tau^+  \tau^-$ and $\mu^+  \mu^-$ have similar branching fractions. 
 Indeed, in both cases  the dominant diagram is the  one with two intermediate $Z$, Fig.~\ref{fig:diagrams}~(c) with practically coincident results.  The next most relevant contribution is different: for $h \to \Upsilon \mu^+ \mu^-$, it comes from  the diagrams with the Higgs coupled to quarks,  Fig.~\ref{fig:diagrams}~(a), while for $h \to \Upsilon \tau^+ \tau^-$ it is  with the Higgs  coupled to leptons,  Fig.~\ref{fig:diagrams}~(b).  The two terms are almost   equal in size in  $h \to \Upsilon \mu^+ \mu^-$ and   $h \to \Upsilon \tau^+ \tau^-$, respectively;   the other diagrams give small  contributions. 

The branching fractions   \eqref{br}  can be compared to those predicted for    $h \to V \gamma$: ${\cal B}(h\to \phi \gamma) = (2.31 \pm 0.11)\times 10^{-6}$ and
${\cal B}(h\to J/\psi \gamma) = (2.95 \pm 0.17)\times 10^{-6}$,  while   ${\cal B}(h\to \Upsilon \gamma)$
 is   ${\cal O}(10^{-9})$ \cite{Koenig:2015pha}.
For the  $h\to V Z$  modes,   ${\cal B}(h \to \phi Z) \simeq  {\cal B}(h\to J/\psi Z) = 2.2\times 10^{-6}$ are  expected in SM  \cite{Isidori:2013cla}.

The  decay distributions in the normalized dilepton mass squared $\hat q^2=q^2/m_h^2$,  Fig.~\ref{fig:distributions}, show that the modes with  final $\mu^+  \mu^-$ pair and those with  $\Upsilon$   are 
 dominated by the  virtual photon and  $Z$ contributions in  Fig.~\ref{fig:diagrams}~(c).  At a
high luminosity facility, such ranges of $\hat q^2$ could be cut in the experimental analysis, to isolate the interferences among the various amplitudes. 
The forward-backward lepton asymmetry is tiny in the whole range of $\hat q^2$. 
For $h \to \phi  \tau^+ \tau^-$ and $h \to J/\psi  \tau^+ \tau^-$ the $\hat q^2$ distributions, in addition to the  $Z$ peak, are enhanced at large dilepton invariant mass,  an effect  of  the diagrams with the Higgs  coupled to the leptons.

\begin{figure}[t!]
\includegraphics[width = .30\textwidth]{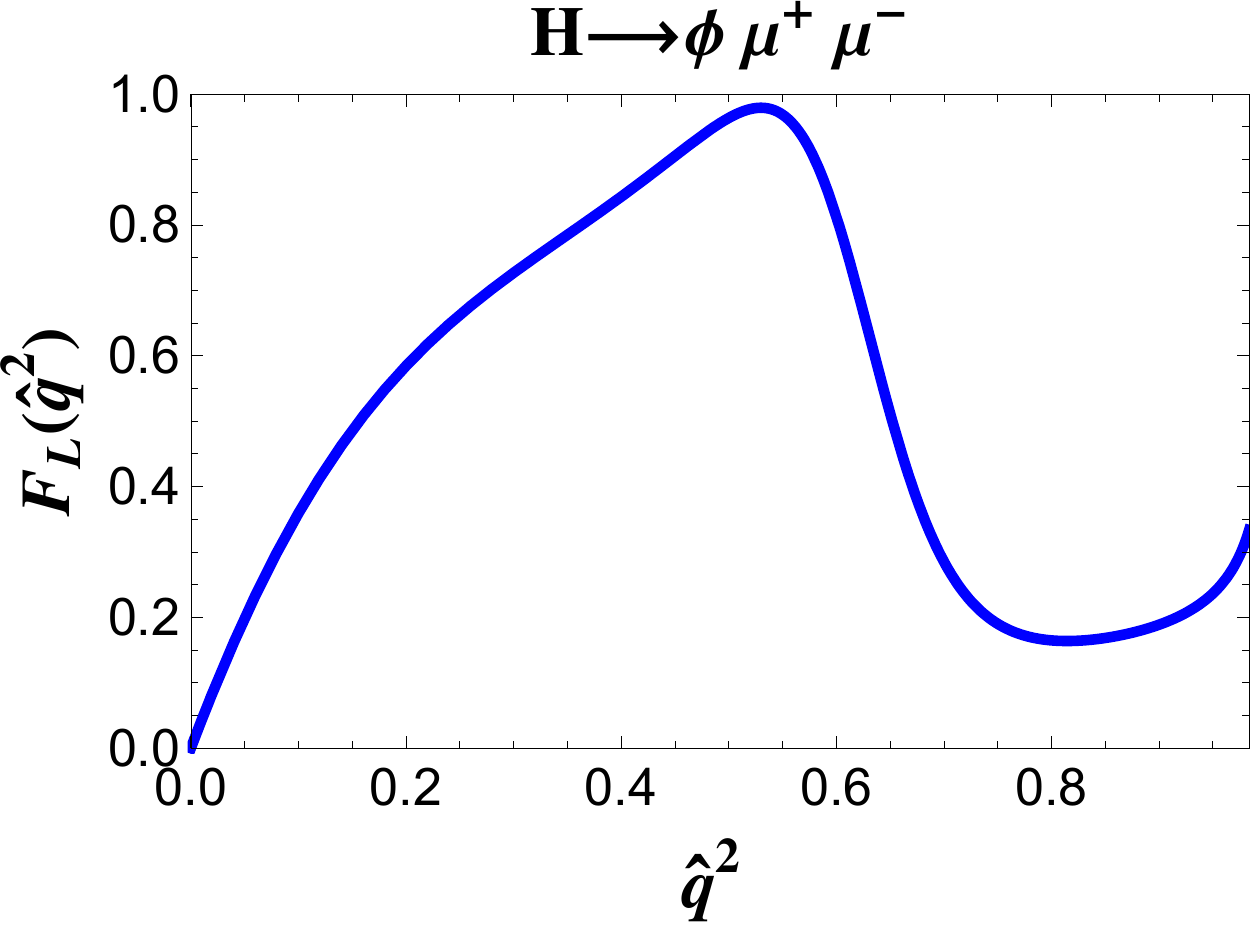}
\includegraphics[width = .30\textwidth]{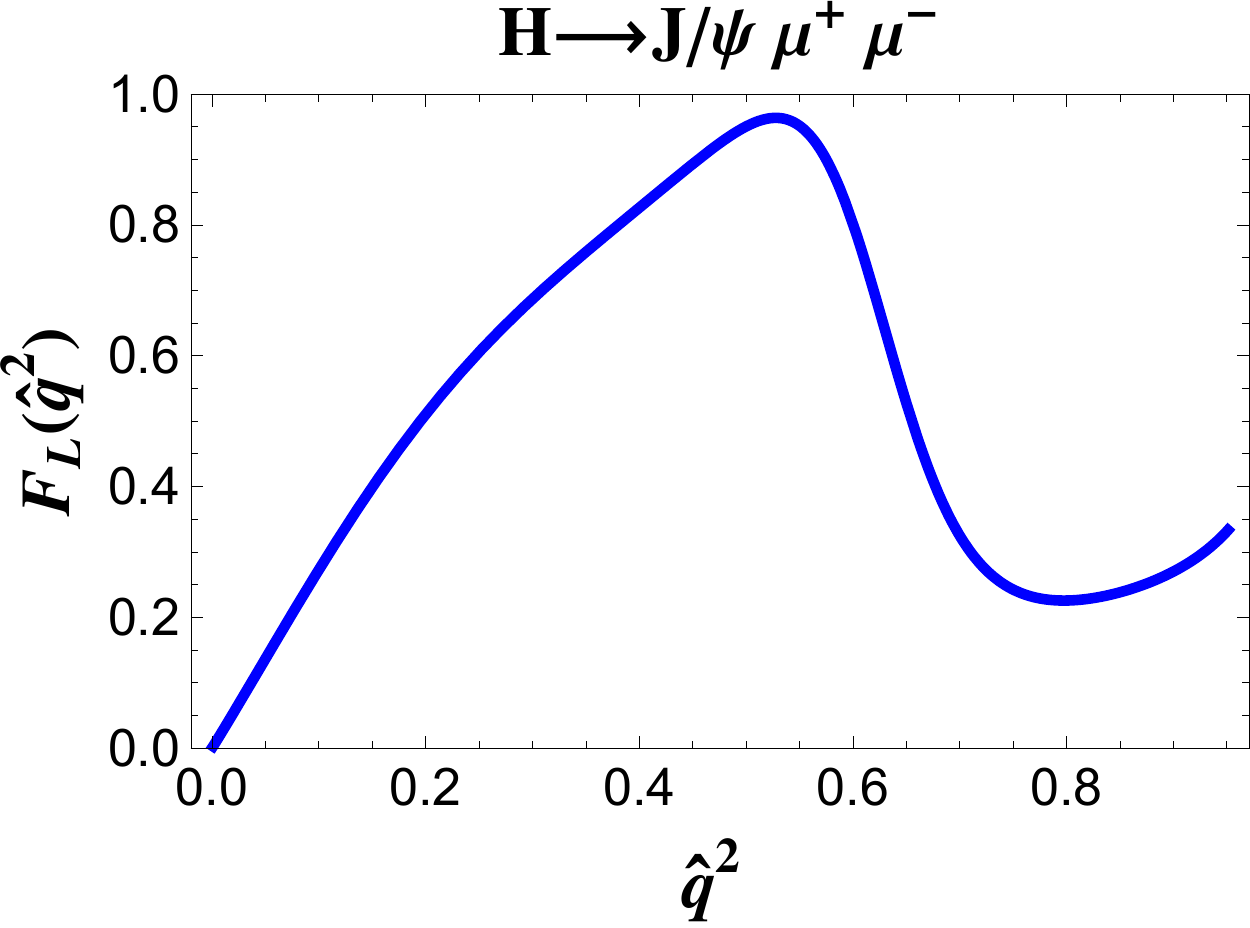}
\includegraphics[width = .30\textwidth]{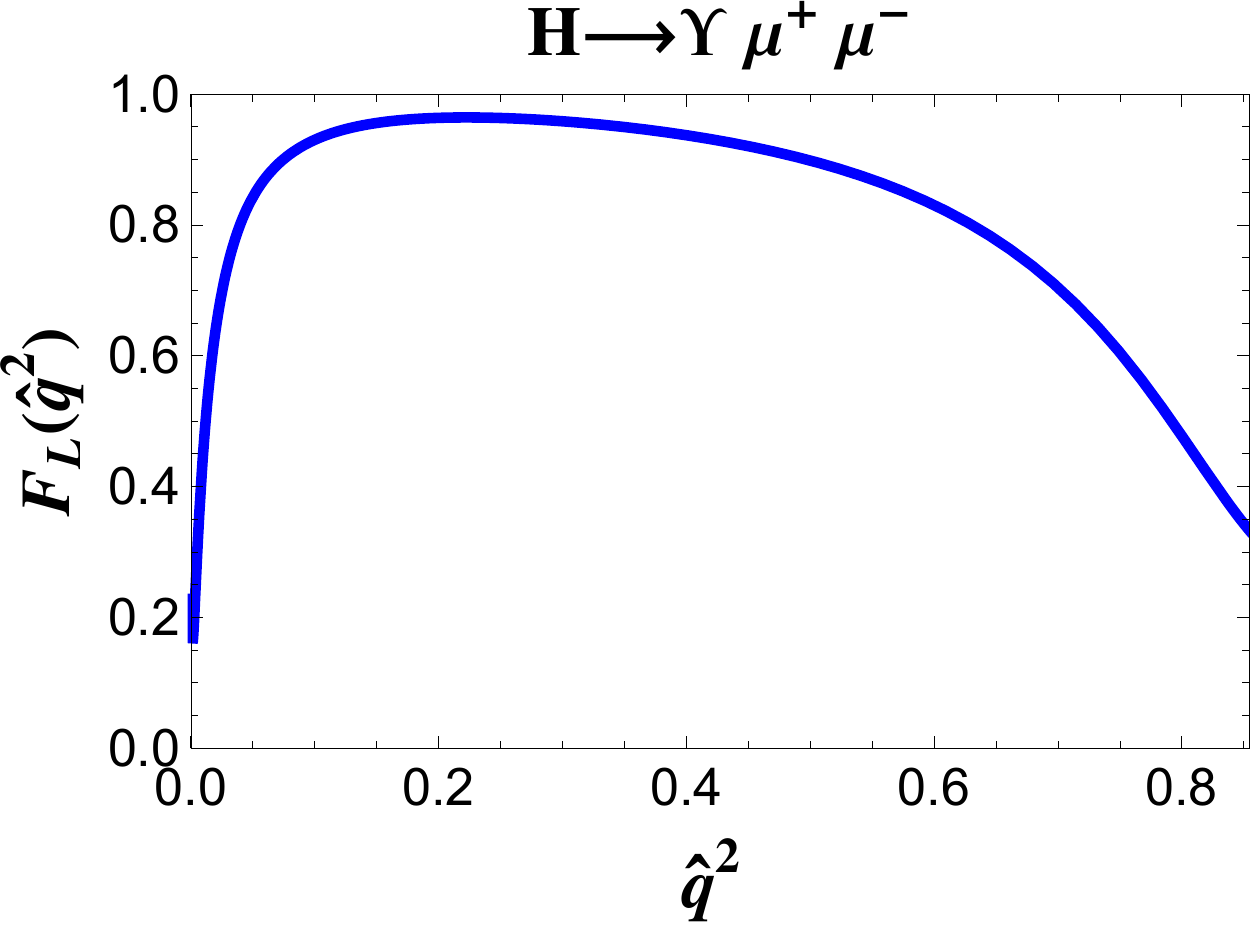}
\\
\\
 \includegraphics[width = .30\textwidth]{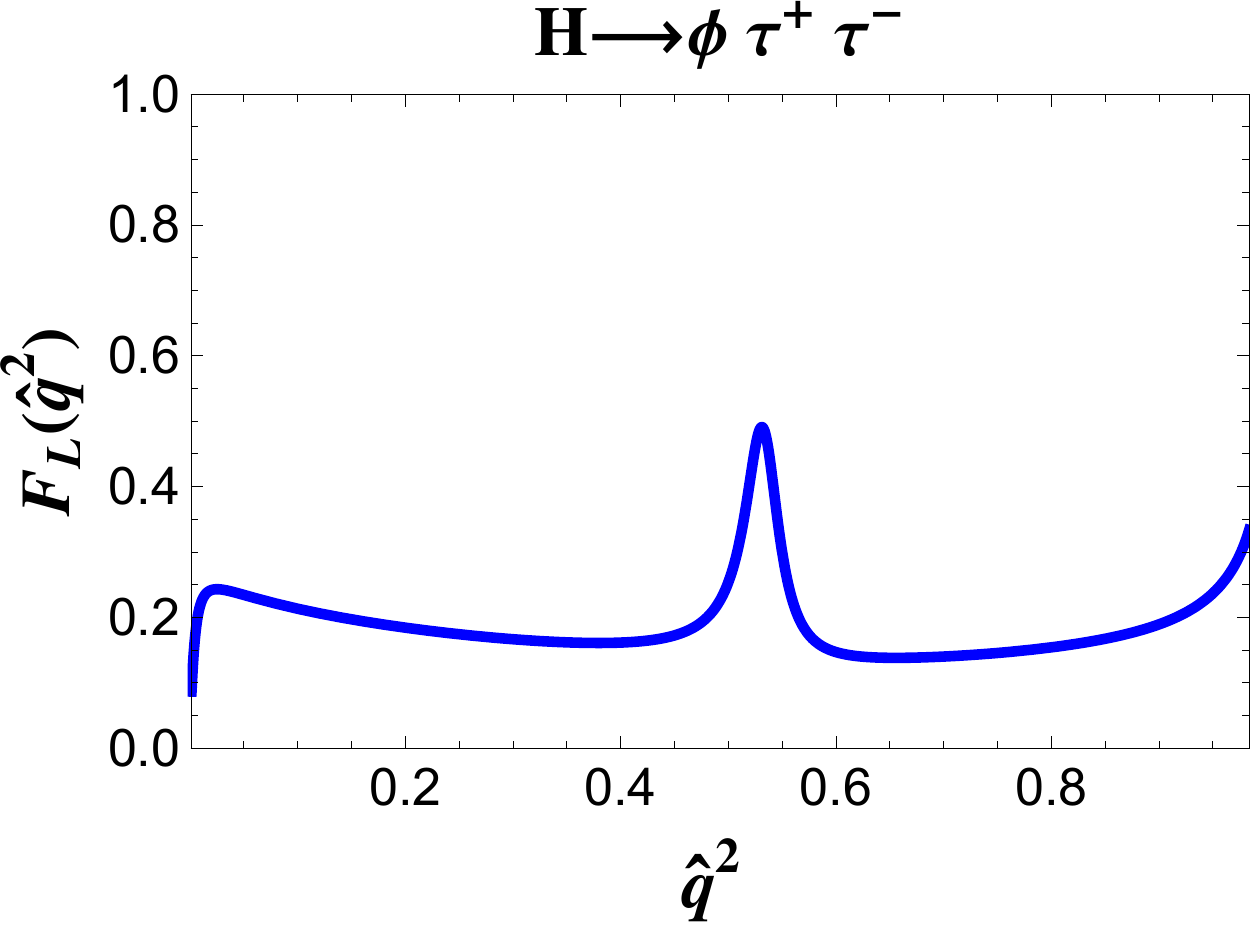}
\includegraphics[width = .30\textwidth]{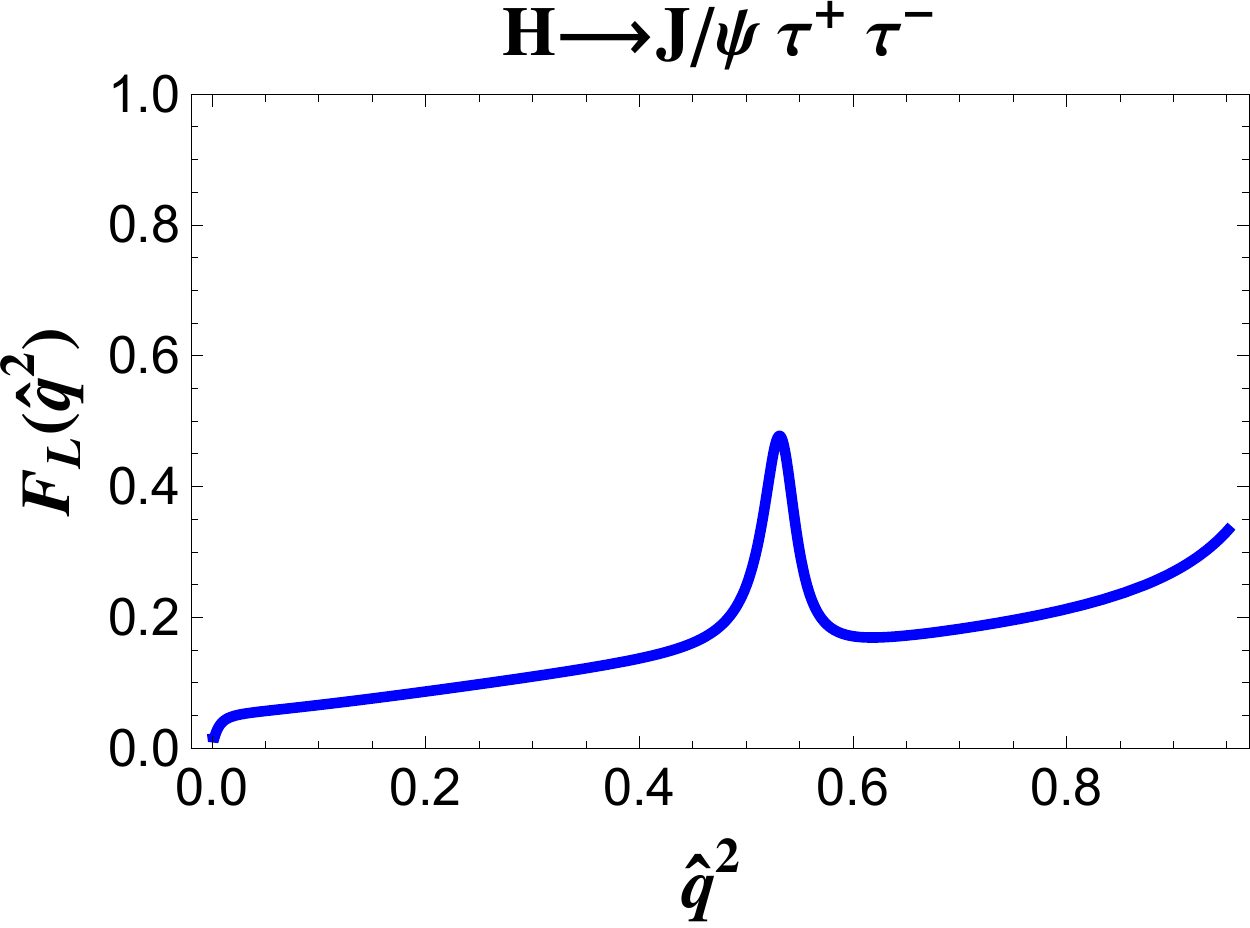}
\includegraphics[width = .30\textwidth]{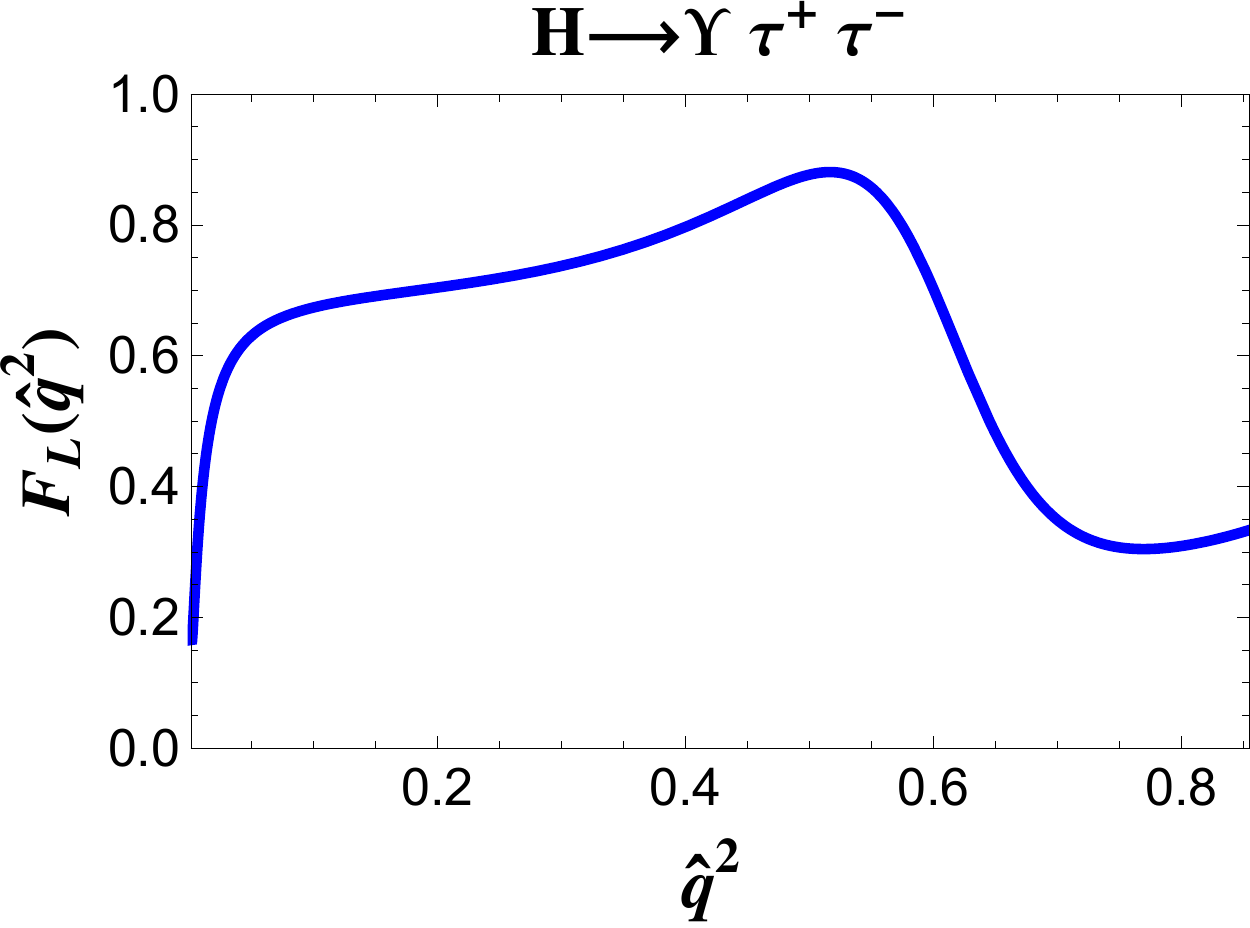}
\caption{Fraction $F_L(\hat q^2)$ of longitudinally polarized meson.}\label{fig:long}
\end{figure}

The  distributions of the fractions of longitudinally polarised vector meson
$\displaystyle F_L(\hat q^2)=\frac{d \Gamma_L (h \to V \ell^+ \ell^-)/d \hat q^2}{d \Gamma(h \to V \ell^+ \ell^-)/d \hat q^2}$ are depicted in Fig.~\ref{fig:long}.
Narrow peaks are found in $\phi \tau^+\tau^-$ and  $J/\psi \tau^+\tau^-$,  in correspondence to the intermediate $Z$, while in the other cases the $\hat q^2$ dependence is milder.  
For modes with  $\mu^+ \mu^-$  one has   $F_L\simeq 1$ at the $Z$ peak, where $Z$ is almost completely longitudinally polarized since both the leptons, in the massless limit, have spins  aligned to the  direction of the motion.

 The $h \to 4 \, \, \ell$ modes, with $\ell=e,\mu$,  have  been analysed  in a kinematical region not far from the intermediate vector resonances, considering only the $h \to Z Z$ contribution,  with the  purpose of determining  the difference in the dilepton spectra in  SM and  in possible extensions~\cite{Gonzalez-Alonso:2014rla,Gonzalez-Alonso:2015bha}. In particular,  a correlation between the channels $h \to 2e2\mu$ and $h \to 4e(4\mu)$ has been recognized  as an observable useful  to identify  the Higgs as a massive excitation of a $SU(2)_L$ doublet,  and to probe the lepton flavour universality of  possible NP contributions \cite{Gonzalez-Alonso:2015bha}. In our analysis we have  included the other diagrams;  furthermore,  we have studied  the modes with  $\tau$ leptons  in  SM.

The $h \to V \nu \bar \nu$  decay widths can be computed  with appropriate changes in  diagrams  in Fig.\ref{fig:diagrams},  predicting
\bea
{\cal B}(h \to \phi \nu \bar \nu) &=& (1.50  \pm 0.075 ) \times 10^{-7} \nn\\
{\cal B}(h \to J/\psi \nu \bar \nu) &=& (1.54 \pm 0.085 ) \times 10^{-7} \label{htonu} \\
{\cal B}(h \to \Upsilon \nu \bar \nu) &=& (1.52 \pm 0.08 ) \times 10^{-6} \,\,\, , \nn 
\eea
with a factor $3$ included to account for the neutrino species.  

\begin{figure}[t!]
\includegraphics[width = .30\textwidth]{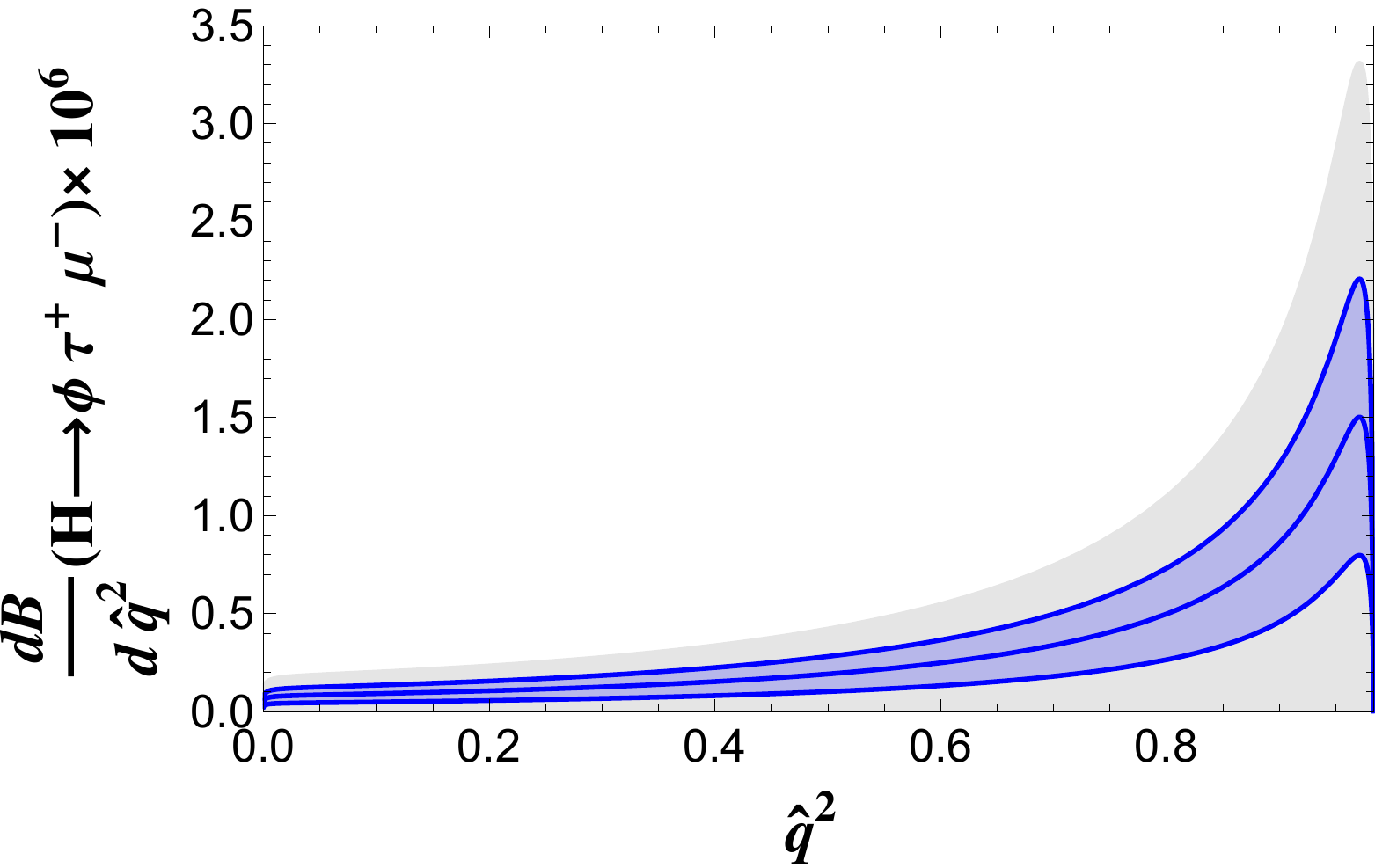}
\includegraphics[width = .30\textwidth]{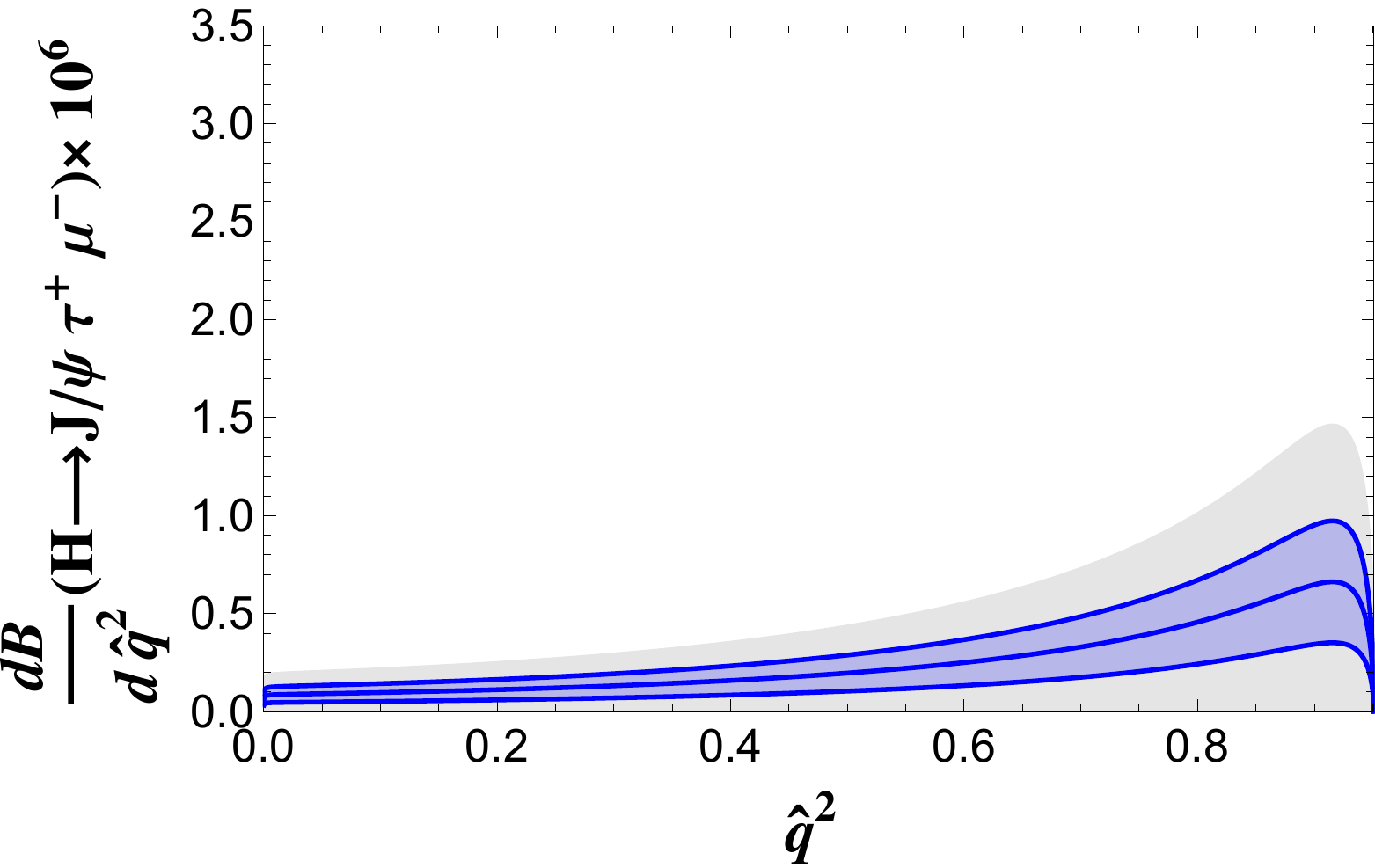}
\includegraphics[width = .30\textwidth]{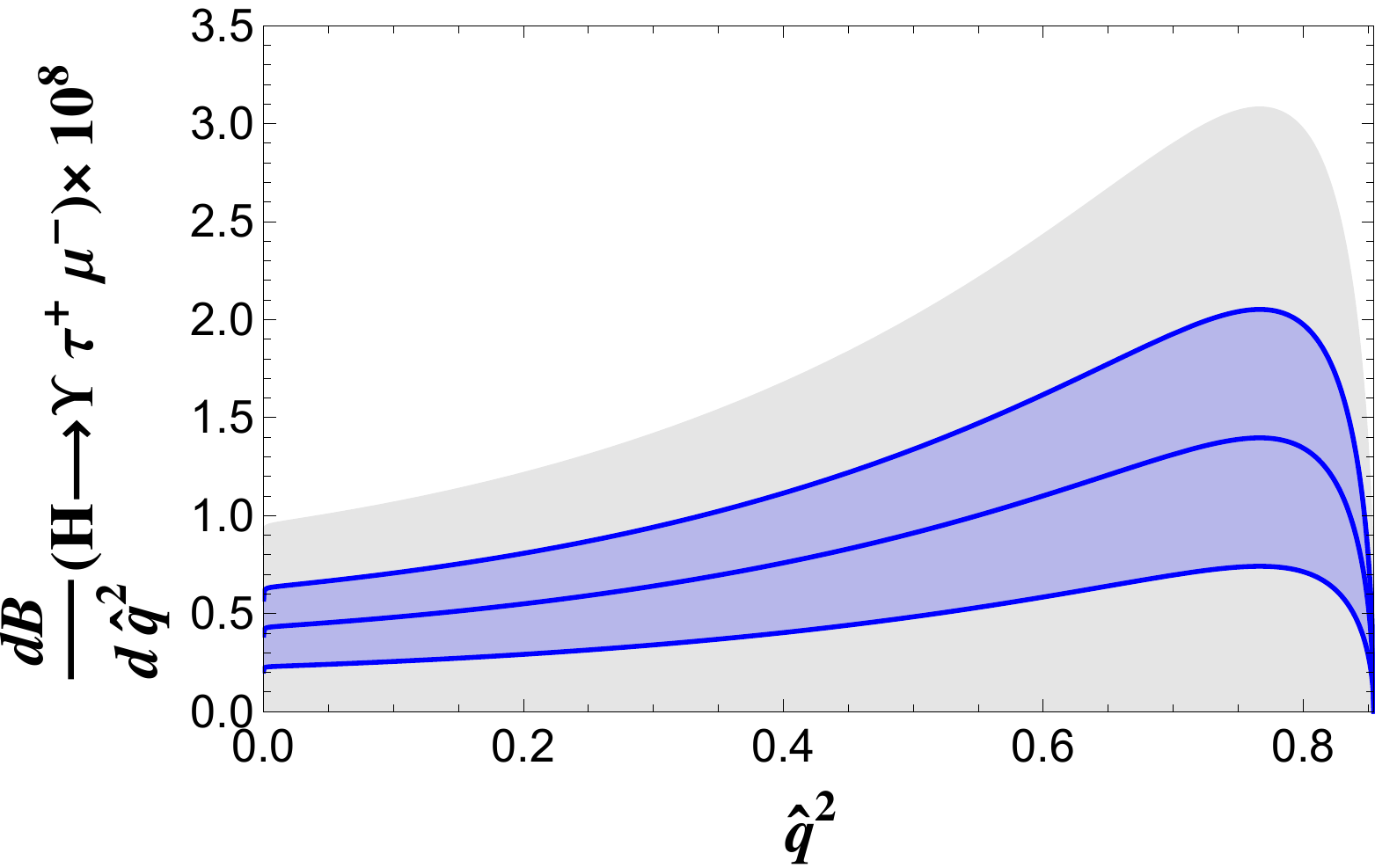}
\caption{Distributions $d {\cal B}(h \to V \tau^+ \mu^-)/d \hat q^2$ obtained in correspondence to the CMS result for $ {\cal B}(h \to  \tau^+ \mu^-)$ \cite{Khachatryan:2015kon}. The light shaded area corresponds to the ATLAS bound in \cite{Aad:2015gha}.}\label{fig:hVtaumu}
\end{figure}

Finally, it is interesting to consider the implications of the LHC studies concerning the lepton flavour violating process $h \to \tau  \mu$ on the exclusive $h \to V \tau  \mu$ processes. The CMS results  correspond to the effective coupling
$\kappa_{h \tau \mu}=(2.6 \pm 0.6)\times 10^{-3}$, considering the uncertainties on ${\cal B}(h \to \tau^+ \mu^-)$ and ${\cal B}(h \to \gamma \gamma)$.
On the other hand, the ATLAS bound corresponds to  $\kappa_{h \tau \mu}< 3.9 \times 10^{-3}$. For these values, the exclusive $h \to V \tau^+ \mu^-$ branching fractions 
and their upper bounds can be computed from the diagrams in Fig.~\ref{fig:diagrams}~(b):
\bea
{\cal B}(h \to \phi \tau^+ \mu^-) &=& (3.2 \pm 1.5) \times 10^{-7} \,\,\, (< 6.9 \times 10^{-7} )\nn \\
{\cal B}(h \to J/\psi \tau^+ \mu^-) &=& (2.4 \pm 1.1) \times 10^{-7} \,\,\, (< 5.2 \times 10^{-7} ) \label{LFV}  \\
{\cal B}(h \to \Upsilon \tau^+ \mu^-) &=& (7.2 \pm 3.4) \times 10^{-9} \,\,\, (< 1.6 \times 10^{-8} ) \,\,\,\, .\nn 
\eea
The  decay distributions  in Fig.~\ref{fig:hVtaumu} have an enhancement at  large  $q^2$.

\vspace*{0.5cm}

In conclusion, for our set of  exclusive $h \to V \ell^+ \ell^-$  decay modes the obtained branching ratios  are in the range  $10^{-8}\div 10^{-6}$ in SM, similar to  
$h\to \phi\gamma$, $h\to \Upsilon \gamma$,  $h\to (\phi,J/\psi) \, Z$. The largest rate is for $h\to \phi \tau^+\tau^-$.
In  the differential $\hat q^2 $ distributions  the  resonant structures at low $q^2$ and at  $q^2=m_Z^2$ are recognized, together with an enhancement at the $q^2$ end-point in 
$h\to\phi\tau^+\tau^-$ and $h\to J/\psi\tau^+\tau^-$. The  rates of  the neutrino modes have  been predicted, 
 the largest one is for $\Upsilon$.  We have also examined the implications of CMS and ATLAS results on  the lepton flavour-changing process $h \to V \tau \mu$. These analyses confirm the role of the exclusive  Higgs boson  decays as  precision tests of the Standard Model and  important probes of physics beyond SM.

\vspace*{0.5cm}
\noindent
One of us (PC) thanks A. Khodjamirian for discussions.

\appendix
\section{Decay amplitudes}\label{amplitudes}
To give the expressions of the amplitudes corresponding to the diagrams in Fig.~\ref{fig:diagrams},
we  define
\be
C_\gamma = 4 \pi \alpha Q_\ell Q_q \,\, , \hspace*{1cm}   
C_Z = \frac{4 \pi \alpha}{s_W^2 c_W^2}  \,,\ee
with $s_W=\sin \theta_W$, $c_W=\cos \theta_W$, and $\theta_W$  the Weinberg angle, and write
the propagators   in Fig.~\ref{fig:diagrams}  in terms of the  functions
\bea
D_1(a,b, {\hat q}^2) &=& a+b {\hat q}^2-ab \, {\hat m}_V^2-{\hat m}_q^2  \,, \nn \\
D_2( {\hat q}^2) &=&  {\hat q}^2- {\hat m}_Z^2+i \, {\hat m}_Z {\hat \Gamma}_Z   \,; \\
D_3({\hat k}) &=& 1-2 n \cdot {\hat k} \,, \nn 
\eea
where $n=(1,{\vec 0})$ and we use  the notation ${\hat x}=\displaystyle{{x}/{m_h}}$, $x$ being a  mass or a momentum.
The 
factorized lepton current has  various Dirac structures. Diagrams with intermediate photons involve  the vector current
\be
V_\ell^\mu  = {\bar \psi}_\ell (k_1) \gamma^\mu \psi_{\bar \ell}(k_2) 
\ee
while, diagrams with intermediate $Z$   also involve 
\bea
A_\ell^\mu & =& {\bar \psi}_\ell (k_1) \gamma^\mu \gamma_5 \psi_{\bar \ell}(k_2)  \, , \nn \\
T_\ell^{\mu \nu}& =&  {\bar \psi}_\ell (k_1) \gamma^\mu \gamma^\nu \psi_{\bar \ell}(k_2) \, , \\
{\tilde T}_\ell^{\mu \nu}& =&  {\bar \psi}_\ell (k_1) \gamma^\mu \gamma^\nu \gamma_5 \psi_{\bar \ell}(k_2) \, . \nn 
\eea
We write  the SM neutral current coupled to  the $Z$ boson as
\be
{\cal L}_\mu=\left( -\frac{ie}{s_W c_W} \right)\left(\Delta_V^f\, {\bar f}\gamma_\mu f + \Delta_A^f\, {\bar f}\gamma_\mu \gamma_5 f \right)
\label{NC}
\ee
where $f$ generically denotes a fermion, and
\be
\Delta_V^f =\frac{1}{2} \left( T_3^f -2 s_W^2 Q^f \right) \, ,  \hspace*{1cm}
\Delta_A^f = -\frac{1}{2}  T_3^f  \, , \ee
with $ T_3^f$  the third component of the weak isospin and $Q^f$ the electric charge of   $f$.
Diagrams  in Fig.~\ref{fig:diagrams}(a) also involve the  integrals over the LCDA of the vector meson $V$:
\bea
I_1 &=&I_1( {\hat q}^2)=\int_0^1\,du \, \phi_\perp^V(u)\left[\frac{1}{D_1(1-u,u,{\hat q}^2)}+\frac{1}{D_1(u, 1-u,{\hat q}^2)} \right]\,,\label{int1}\\
I_2 &=&I_1( {\hat q}^2)=\int_0^1\,du \, \phi_\perp^V(u)\left[\frac{u}{D_1(1-u,u,{\hat q}^2)}+\frac{1-u}{D_1(u, 1-u,{\hat q}^2)} \right]\,.\label{int2}
\eea
With these  definitions, the amplitudes in Fig.~\ref{fig:diagrams} can be written.  We report the various expressions  in  correspondence with the diagrams in 
Fig.~\ref{fig:diagrams}(a), (b) and (c),  considering separately the intermediate photon and $Z$ contributions.
\begin{itemize}
\item Fig. \ref{fig:diagrams}(a), intermediate $\gamma$:
\be
A_{(a)}^\gamma=C_{(a)}^\gamma m_h \epsilon_V^{*\alpha} V_{\ell \, \mu} \, \left\{[n_\alpha {\hat p}_V^\mu- g_\alpha^\mu (n \cdot {\hat p}_V)] I_1-g^\mu_\alpha {\hat m}_V^2 I_2\right\}
\ee
with
\be
C_{(a)}^\gamma=\frac{1}{m_h^2} \frac{\hat m_q}{v} C_\gamma  f_V^\perp  \frac{1}{\hat q^2}  \,\,\, .
\ee
\item  Fig.~\ref{fig:diagrams}(a),  intermediate $Z$:
\be
A_{(a)}^Z=C_{(a)}^Z \epsilon_V^{*\alpha}
\left[\Delta_V^\ell \, V_{\ell \, \mu}+ \Delta_A^\ell \, A_{\ell \, \mu}\right]\, \left( g^{\mu \alpha} p_V^\sigma -g^{\alpha \sigma} p_V^\mu \right)\,\left[ n_\sigma I_1-{\hat p}_{V \sigma} I_2 \right]
\ee
with
\be
C_{(a)}^Z=-\frac{1}{m_h^2} \frac{\hat m_q}{v} C_Z \frac{1}{D_2({\hat q}^2)} f_V^\perp \,\Delta_V^q\,\,\, .
\ee
\item  Fig. \ref{fig:diagrams}(b),  intermediate $\gamma$:
\be
A_{(b)}^\gamma=C_{(b)}^\gamma \epsilon_V^{*\alpha} n^\mu \left[-\frac{1}{D_3({\hat k}_1)}\, T_{\ell \, \mu \alpha}+\frac{1}{D_3({\hat k}_2)}\, T_{\ell \,  \alpha \mu} \right]
\ee
with
\be
C_{(b)}^\gamma = \frac{1}{m_h^2} \frac{\hat m_\ell}{v} C_\gamma \frac{f_V m_V}{\hat m_V^2} \,\,\, .
\ee
\item Fig. \ref{fig:diagrams}(b), intermediate $Z$:
\be
A_{(b)}^Z=C_{(b)}^Z \epsilon^*_{V \alpha} n_\mu \left\{-\frac{1}{D_3({\hat k}_1)} \left[\Delta_V^\ell T_\ell^{\mu \alpha}+\Delta_A^\ell {\tilde T}_\ell^{\mu \alpha} \right] +\frac{1}{D_3({\hat k}_2)} \left[\Delta_V^\ell T_\ell^{ \alpha \mu}-\Delta_A^\ell {\tilde T}_\ell^{ \alpha \mu} \right] \right\}
\ee
with
\be
C_{(b)}^Z=\frac{1}{m_h^2}\frac{\hat m_\ell}{v} C_Z \frac{\Delta_V^q}{D_2({\hat m}_V^2)}f_V m_V   \,\,\, .
\ee
\item  Fig. \ref{fig:diagrams}(c), two intermediate photons:
\be
A_{(c)}^{\gamma \gamma}=C_{(c)}^{\gamma \gamma} \epsilon_V^{*\alpha}[ g_{\alpha \mu} (q \cdot p_V)-m_h^2 n_\alpha n_\mu] \, V_{\ell}^{\mu}
\ee
with
\be
C_{(c)}^{\gamma \gamma}=\frac{1}{m_h^4}\frac{\alpha}{\pi v} C_{\gamma \gamma} \, C_\gamma \frac{f_V m_V}{\hat m_V^2} \frac{1}{{\hat q}^2} \,\,\, .
\ee
\item Fig. \ref{fig:diagrams}(c),  two intermediate $Z$:
\be
A_{(c)}^{ZZ}=C_{(c)}^{ZZ}\epsilon^*_{V \alpha} \left(\Delta_V^\ell \,V_\ell^\alpha + \Delta_A^\ell \,A_\ell^\alpha \right)
\ee
with
\be
C_{(c)}^{ZZ}=\frac{1}{m_h^2} \frac{2 {\hat m}_Z^2}{v}C_Z \frac{1}{D_2({\hat q}^2)}\frac{1} {D_2({\hat m}^2_V)} \Delta_V^q f_V m_V  \,\,\, .
\ee
\item  Fig. \ref{fig:diagrams}(c),  intermediate  $\gamma \, Z$, with  $\gamma$ converting to leptons:
\be
A_{(c)}^{\gamma Z}=C_{(c)}^{\gamma Z}\epsilon_V^{*\alpha}[ g_{\alpha \mu} (q \cdot p_V)-m_h^2 n_\alpha n_\mu] \, V_{\ell}^ {\mu}
\ee
with
\be
C_{(c)}^{\gamma Z}=\frac{1}{m_h^4}\frac{\alpha}{\pi v}C_{\gamma Z} \frac{4 \pi \alpha Q_\ell}{s_W c_W} \frac{1}{{\hat q}^2} \frac{\Delta_V^q}{D_2({\hat m}^2_V)} f_V m_V \,\,\, .
\ee
\item  Fig. \ref{fig:diagrams}(c),  intermediate  $Z \, \gamma$, with $Z$ converting to leptons:
\be
A_{(c)}^{Z \gamma }=C_{(c)}^{Z \gamma }\epsilon_V^{*\alpha}[ g_{\alpha \mu} (q \cdot p_V)-m_h^2 n_\alpha n_\mu] \, \left(\Delta_V^\ell \,V_\ell^\mu + \Delta_A^\ell \,A_\ell^\mu \right) 
\ee
with
\be
C_{(c)}^{Z \gamma }=\frac{1}{m_h^4}\frac{\alpha}{\pi v}C_{\gamma Z} \frac{4 \pi \alpha Q_q}{s_W c_W}\frac{1}{{\hat m}^2_V} \frac{1}{D_2({\hat q}^2)} f_V m_V \,\,\, .
\ee
\end{itemize}
The effective couplings $C_{\gamma \gamma} $ and $C_{\gamma Z}$ are defined through Eq.~\ref{effective}.

\bibliographystyle{JHEP}
\bibliography{ref-higgs}
\end{document}